\documentclass{aastex61}
\usepackage{amsmath}
\usepackage{color}
\usepackage{float}
\usepackage{multirow}
\usepackage{upgreek}
\usepackage{natbib}

\newcommand{\kepler}{{\it Kepler}}
\newcommand{\tess}{{\it TESS}}
\shorttitle{Binary Systems Search with \tess}
\shortauthors{Otani et al.}
\graphicspath{{./}{figures/}}

\accepted{July 29, 2025}
\submitjournal{AJ}

\begin{document}
\title{A Search for Subdwarf B Binaries using Pulsation Timing from {\it TESS}} 

\correspondingauthor{Tomomi Otani}
\email{otanit@erau.edu}

\author[0000-0002-0786-7307]{Tomomi Otani} 
\affiliation{Department of Physical Sciences, Embry-Riddle Aeronautical University, 1 Aerospace Blvd, Daytona Beach, FL 32114, USA}

\author{A. S. Baran}
\affiliation{Physics Astronomy and Material Science Department, Missouri State University, 901 S National Ave, Springfield, MO 65897, USA}
\affiliation{ARDASTELLA Research Group}

\author{Lindsay C. Spence}
\affiliation{Department of Physical Sciences, Embry-Riddle Aeronautical University, 1 Aerospace Blvd, Daytona Beach, FL 32114, USA}

\author[0000-0002-5775-2866]{Ted von Hippel}
\affiliation{Department of Physical Sciences, Embry-Riddle Aeronautical University, 1 Aerospace Blvd, Daytona Beach, FL 32114, USA}

\author{E. Lynum-Lozano}
\affiliation{Department of Physical Sciences, Embry-Riddle Aeronautical University, 1 Aerospace Blvd, Daytona Beach, FL 32114, USA}

\author{Julia. M. Clark}
\affiliation{Department of Physical Sciences, Embry-Riddle Aeronautical University, 1 Aerospace Blvd, Daytona Beach, FL 32114, USA}


\begin{abstract}

Hot subdwarf B (sdB) stars are post-main-sequence stars of high temperature and gravity. Approximately 30\% of sdBs  exhibit stable pressure and/or gravity-mode pulsations, which can be used via the timing method to test for companion stars and determine their orbital solutions. We used short cadence data from the Transiting Exoplanet Survey Satellite (TESS) to search for previously undiscovered companions to sdBs. In this paper, we focus on searching for companions with orbital periods shorter than 13.5\,d which are detectable within one sector of \tess\ data (about 27\,d). The timing method requires that we derive pulsation frequencies in subsets of data significantly shorter than the periods we are searching for, which we set at 0.5 to 1.5\,d.  We investigated ten sdB stars with previously detected p-mode pulsations for which at least one p-mode pulsation remains detectable with a signal-to-noise ratio (S/N) $>$ 4 within data subsets of duration 0.5 - 1.5\,d. We find that two (TIC\,202354658 and TIC\,69298924) of these ten sdB stars likely have white dwarf companions and set limits on companion masses for the other eight sdB stars.
\end{abstract}

\section{Introduction} \label{sec:intro}
Subdwarf B (sdB) stars are extreme horizontal branch (EHB) stars of high temperature ($T_{\rm eff}=22\,000$ to $32\,000$ K) and gravity (log g = 5.0 to 6.2). They are core helium-burning objects, found in both the Galactic disk and halo \citep{saffer94}.  An sdB progenitor, while on the red giant branch, loses most of its hydrogen envelope prior to the helium flash  \citep{dcruz96,mengel76}.  Because of degenerate core evolution, their masses are narrowly confined to approximately 0.5\,M$_{\sun}$ \citep{ostrowski21}.  After the red giant branch, the star burns helium in a shell for up to 10$^{8}$\,yr, however nuclear burning will not be sustained in the hydrogen envelope, and therefore the star cannot ascend the asymptotic giant branch after helium core exhaustion \citep {heber86} and instead will directly evolve to become a low-mass helium white dwarf (WD) \citep {bergeron97}.

The most studied sdB evolution models are Binary Population Synthesis (BPS) models constructed by \citet{han02,han03} that included three evolutionary channels: [1] common envelope (CE) ejection, which can produce sdB binary systems with short orbital periods, [2] Roche-lobe overflow (RLOF), which can produce sdB binary systems with long orbital periods, and [3] white dwarf mergers, which can produce single sdB stars. For the CE ejection channel the secondary component is either a main sequence (MS) star or a WD, with a typical orbital period between 0.1 and 20\,d. Many observational studies have focused on these short orbital period systems \citep[e.g.][]{copperwheat11}. To date, the orbital parameters for approximately 140 short orbital period binaries have been derived, and more than half of these have orbital periods shorter than 0.6\,d \citep{ge22}. As seen in Fig.\,3 of \citet{ge22}, radial velocity (RV) surveys to find sdB binaries with orbital periods longer than 10\,d are infrequent because RV searches are inefficient in this period range. Only 11 sdB + WD binary systems with orbital periods longer than five days have been identified. Searching for binaries with orbital periods longer than typical short orbital period binary systems provides valuable insight into the transition from detached post-CE configurations to RLOF, shedding light on angular momentum loss mechanisms and the stability of mass transfer \citep{ivanova13}.

Approximately 30\% of sdB stars show detectable pulsations. The existence of sdB pulsators (sdBV, where "V" stands for variable) was theoretically predicted by \citet{charpinet97}.  Independently, \citet{kilkenny97} observationally discovered the very first short period sdBV star, \objectname{EC\,14026-2647}. These stars are pressure (p) mode pulsators, with pulsations driven by internal pressure fluctuations \citep{charpinet00}. The very first long-period gravity (g) mode sdBV star, \objectname{PG\,1716+426}, was reported by \cite{green03}, in which buoyancy provides the restoring force. Some sdB stars have been discovered to exhibit both p- and g-mode pulsations. These objects are referred to as hybrid pulsators \citep{baran05}. 
      
Although amplitudes were found to vary (e.g. \citealp{kilkenny10}), the pulsation periods of sdBV stars are usually stable \citep{ostensen01}, and therefore good chronometers. As is well known, a star's position in space may show cyclic variations due to a gravitational perturbation of a companion. From an observer's point of view, the light from a pulsating star is periodically delayed when it is on the far side of its orbit and advanced on the near side. The orbital solution of the binary system can be obtained by measuring changes in the pulsation arrival times, which is also called light time delay.  This technique has been used to search for additional components to pulsating stars and to derive orbital parameters of those systems.
      
The presence of an M dwarf or WD companion to the sdBV star CS\,1246, suggested by this technique, was confirmed by RV measurements \citep{barlow11}. \citet{otani18} used this technique to obtain an orbital solution of the previously known sdB + main sequence binary system, EC\,20117-4014, while \citet{otani22} detected one of the companions to the sdB triple star system candidate, AQ\,Col. 

Continuous long-coverage photometric data obtained from space telescopes such as \kepler\, and \tess\, allow us to efficiently search for companions using this method. Recently, a short orbital period sdB plus WD binary system, BPM\,36430, was detected using this method with \tess\ data and confirmed via radial velocities \citep{smith22}.  

This work is a continuation of our search for additional components to sdB stars \citep{otani18, otani22} with this pulsation timing method using TESS photometry. Section~\ref{sec:obser} outlines the targets and their data. Section~\ref{sec:analysis} explains our analysis. Section~\ref{sec:results} presents our results and discussion. Our conclusions and future work are summarized in Section~\ref{sec:conc}.



\section{Photometrtic Data} \label{sec:obser}

\tess\ was launched in 2018 and provides extended time-series photometry for millions of objects across the entire sky \citep{ricker14}. We have chosen 10 hot subdwarfs (see Table\,1) with previously reported p-mode pulsations, observed by \tess\ during Cycle 3 and/or 4 using 20\,sec cadence (USC) data to properly sample minute-long p-modes. These stars have either been previously identified as single stars or flagged as possible binary systems for which orbital elements (orbital period, a\,$\sin{\rm i}$, eccentricity, and the mass function) have not been derived.
We accessed the time-series data from the Mikulski Archive for Space Telescopes (MAST) and used \uppercase{PDCSAP\_FLUX}, which is simple aperture photometry (\uppercase{SAP\_FLUX}) corrected for on-board systematics, as well as the \tess\ Barycentric Julian Date (TBJD). 


\begin{table}
\centering
\caption{Pulsating sdB stars investigated in this work}
\label{tab:SCfreq}
\begin{tabular}{llccllc}
\hline\hline
\multicolumn{1}{c}{\multirow{2}{*}{ID}} & \multicolumn{1}{c}{\multirow{2}{*}{Other name}} & \multicolumn{1}{c}{RA} & \multicolumn{1}{c}{DEC} & \multicolumn{1}{c}{G$_{\rm BP}$-G$_{\rm RP}$} & \multicolumn{1}{c}{G} &\multicolumn{1}{c}{\multirow{2}{*}{\tess\ Sector}}\\
& & \multicolumn{1}{c}{[deg]} & \multicolumn{1}{c}{[deg]} & \multicolumn{1}{c}{[mag]} & \multicolumn{1}{c}{[mag]} \\
\hline\hline
TIC\,47377536 & PG\,1047+003 & 162.5 & -0.0 & -0.4647(42) & 13.4190(10) & 35, 46, 72\\ 
TIC\,62381958 & EC\,01541-1409 & 29.1 & -13.9 & -0.2618(37) & 12.2460(10) & 30 \\ 
TIC\,62483415 & EC\,22221-3152 & 336.2 & -31.6 & -0.4852(40) & 13.6640(9) & 28, 68\\ 
TIC\,69298924 & J08069+1527 & 121.7 & 15.5 & -0.4027(49) & 14.7526(12) & 44--46, 71--72\\ 
TIC\,136975077 & KPD\,2109+4401 & 317.9 & 44.2 & -0.3459(23) & 13.3435(7) & 55--56, 75--76, 82--83\\ 
TIC\,165312944 & PG\,1219+534 & 185.4 & 53.1 & -0.4879(18) & 13.1655(5) & 48--49, 75--76\\ 
TIC\,169285097 & J2344-3427 & 356.1 & -34.5 & -0.4302(15) & 10.9142(7) & 29, 69\\ 
TIC\,202354658 & 2M\,1545+5955 & 236.3 & 59.9 & -0.4399(31) & 14.4089(11) & 48--51, 58, 75--78, 82, 85\\ 
TIC\,248949857 & PB\,8783 & 20.9 & -5.1 & 0.2218(13) & 12.2189(4) & 30\\ 
TIC\,396954061 & 2M\,0415+0154 & 64.0 & 1.9 & -0.2959(31) & 14.0278(7) & 32\\ 
\hline\hline
\end{tabular}
\end{table}

\section{Analysis} \label{sec:analysis}
A flux variation caused by pulsations can serve as an accurate clock. A specific phase of the flux variation can be compared with predictions. Traditionally, times of flux maxima for pulsating stars are compared from a linear ephemeris, which describes times of consecutive flux maxima. This technique is known as Observed\,--\,Calculated (O$-$C) analysis. \citet{sterken05} discussed multiple possible solutions, with some being only corrections to either a period or reference epoch adopted in the linear ephemeris, while parabolic and sinusoidal trends mean a physical interpretation, {\it i.e.} evolutionary change and an additional companion, respectively. The O$-$C analysis is useful for a variable star that has a single periodic variation; however, it is not suitable for stars with multiperiodic variations.  \citet{murphy14} discussed the phase modulation (PM) method, which is a complementary method of the traditional O$-$C method, by which the phase modulation is extracted from the Fourier analysis of the time-shifted light curve due to the binary orbital motion.  Since most of the sdB pulsators are multiperiodic, this method was used to obtain the light time delay in our analysis. In this work we search for periodic behavior in the light time delay indicative of an additional companion to the pulsating star. In general, the orbit of the companion may be eccentric hence the light time delay variation trend will be described with Bessel functions \citep[see][]{otani22} rather than a sine function. The Bessel functions include the orbital period $P$, a$_{\rm sdB}\sin{i}$, orbital eccentricity $e$, and the argument of periapsis $\varpi$. In cases with negligible orbit eccentricity, one can use a sine function, which assumes $e$\,=\,0, while $\varpi$ becomes undefined. From a$_{\rm sdB}\sin{i}$, we can also derive the mass function \citep[details can be found in Equation (22) of][]{otani22}, and the RV amplitude, $K$, of the pulsating star. The RV and light time delay variation are discussed and used in {\it e.g.} \citet{telting12}.

To obtain the light time delay variations, we split all photometry into 0.5\,d data subsets (1.5\,d for TIC\,69298924), and check which pulsation frequencies are still significant in those individual data subsets. Then, we derived times of flux maxima of the significant pulsation frequencies and compared these with predictions calculated from the ephemerides. A reference epoch was adopted as the timestamp of the first data point, while a period is the average value fitted across all data for that system.

Note that two closely spaced pulsation frequencies, {\it e.g.} rotationally split modes, also called multiplets, can be unresolved in data subsets and cause sinusoidal variations in the light time delay diagram, even though the variation is not because of an 'actual' light time delay.  This can be identified because in such a case, the pulsation amplitudes are also periodically modulated.

When there is a companion to a pulsating star, the pulsation frequency is surrounded by two additional frequencies, which altogether looks like a triplet, and can be misinterpreted as rotationally split modes.
Yet these three frequencies show a different behavior. There is a sinusoidal trend in the light time delay diagram, however the amplitude of the analyzed frequency is unmodulated. In addition, the amplitudes of the two side frequencies should obey a relation cited in Sections\,2.4.2 and 2.4.3 of \citet{shibahashi12}. Following these approaches, our conclusion on the existence of additional companions is based on whether the amplitude shows in-phase variability with times of flux maxima, as well as the correct amplitude prediction of the side frequencies.



\section{Results and Discussion} \label{sec:results}    

\begin{figure*} [ht]
\includegraphics[width=\hsize,height=23cm]{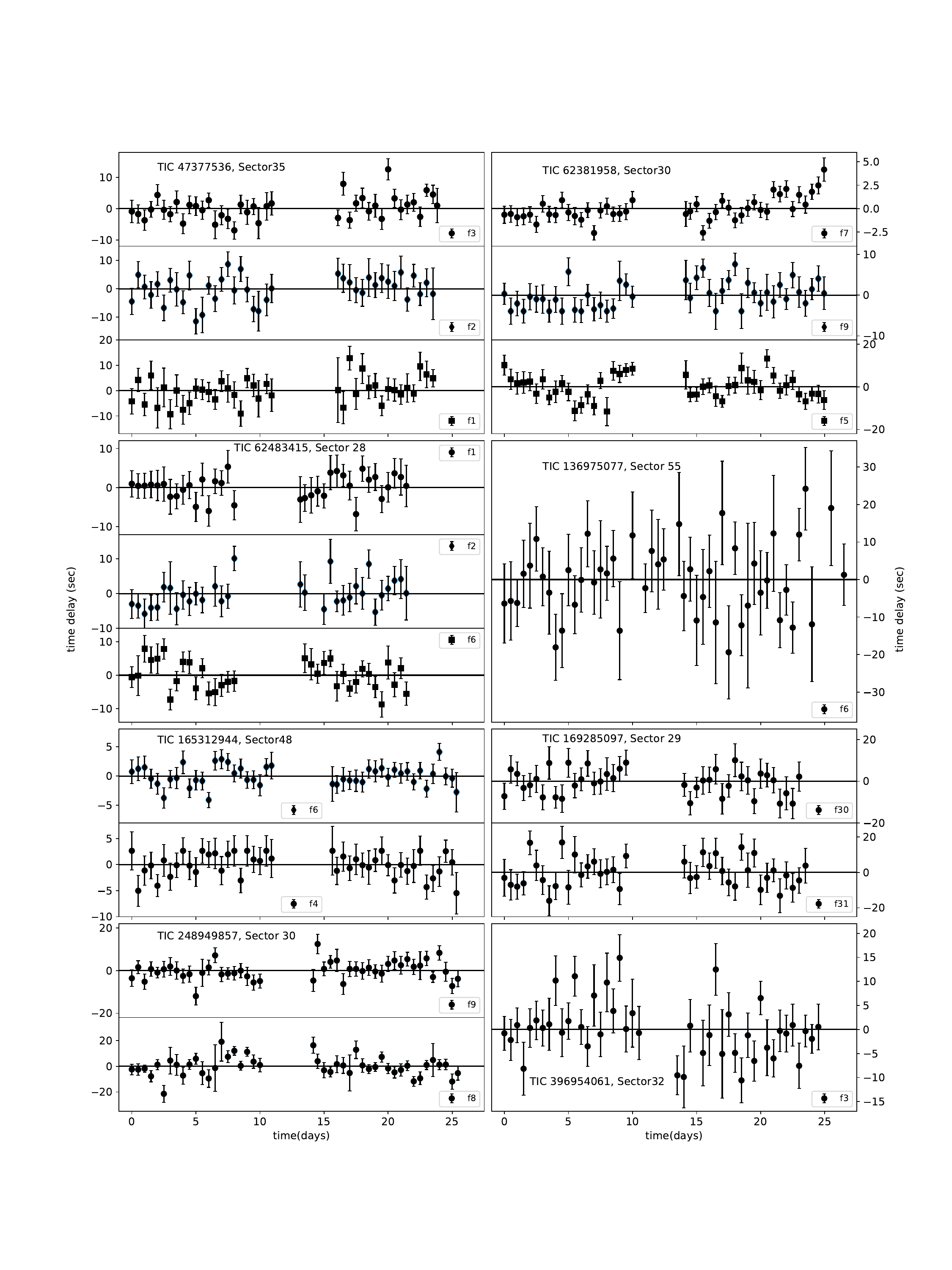}
\caption{Pulsation timing analysis of targets with no observed periodic variations. Only the first sector in which each target was observed with a 20-second cadence is shown. When more than one pulsation frequency is used for the analysis, the pulsation with the largest amplitude is shown on top.} 
\label{fig:OC1}
\end{figure*}

TIC\,47377536 (PG\,1047+003, UY\,Sex, Gaia\,DR2\,3806303066866089216) was found as a variable sdB star independently by \citet{billeres97} and \citet{odonoghue98}, who indicated that its optical spectrum is consistent with that of a single sdB star.  According to \citet{kilkenny02} the effective temperature and the surface gravities are T$_{\rm eff}$\,=\,34\,700\,K and $\log$\,g\,=\,5.8\,$\pm$\,0.15. \citet{reed20} analyzed photometric data obtained during Campaign\,14 of the K2 mission, and detected 97 frequencies. \citet{baran23} analyzed \tess\ USC data collected in Sectors\,35 and 46 and detected frequencies with high amplitudes in both sectors.
We obtained the light time delay diagram on these \tess\ USC data by splitting them into 0.5 d data subsets, using only frequencies that were detected above S/N\,=\,4 within these data subsets, which for this star was f$_1$, f$_2$, and f$_3$ in Sector\,35 and f$_2$ and f$_6$ in Sector\,64. The light time delay diagram is shown in Figure\,\ref{fig:OC1}. We found no significant periodic variations. 

These timing residuals show no variation with an amplitude higher than 5.41\,sec, defined as four times the average noise level in the amplitude spectrum calculated from the light time delay diagram data points. Only the result of the light time delay for f$_2$ were used since f$_2$ provides the lowest average noise level. We used this value to estimate the upper limit of the possible companion that could still exist. Using the amplitude of the light time delay diagram (the amplitude in Equation (6) of Otani et al.2022), $\tau = \frac{1}{c} a_{sdB} \sin{i}$, the mass function can be written as:

\begin{equation}
f = \frac{(M_{comp} \sin{i})^3}{(M_{sdB}+M_{comp})^2} = \frac{4\pi^2 c^3 \tau^3}{GP_{orb}^2}.
\label{eq:1}
\end{equation}

\begin{figure}[] 
\plotone{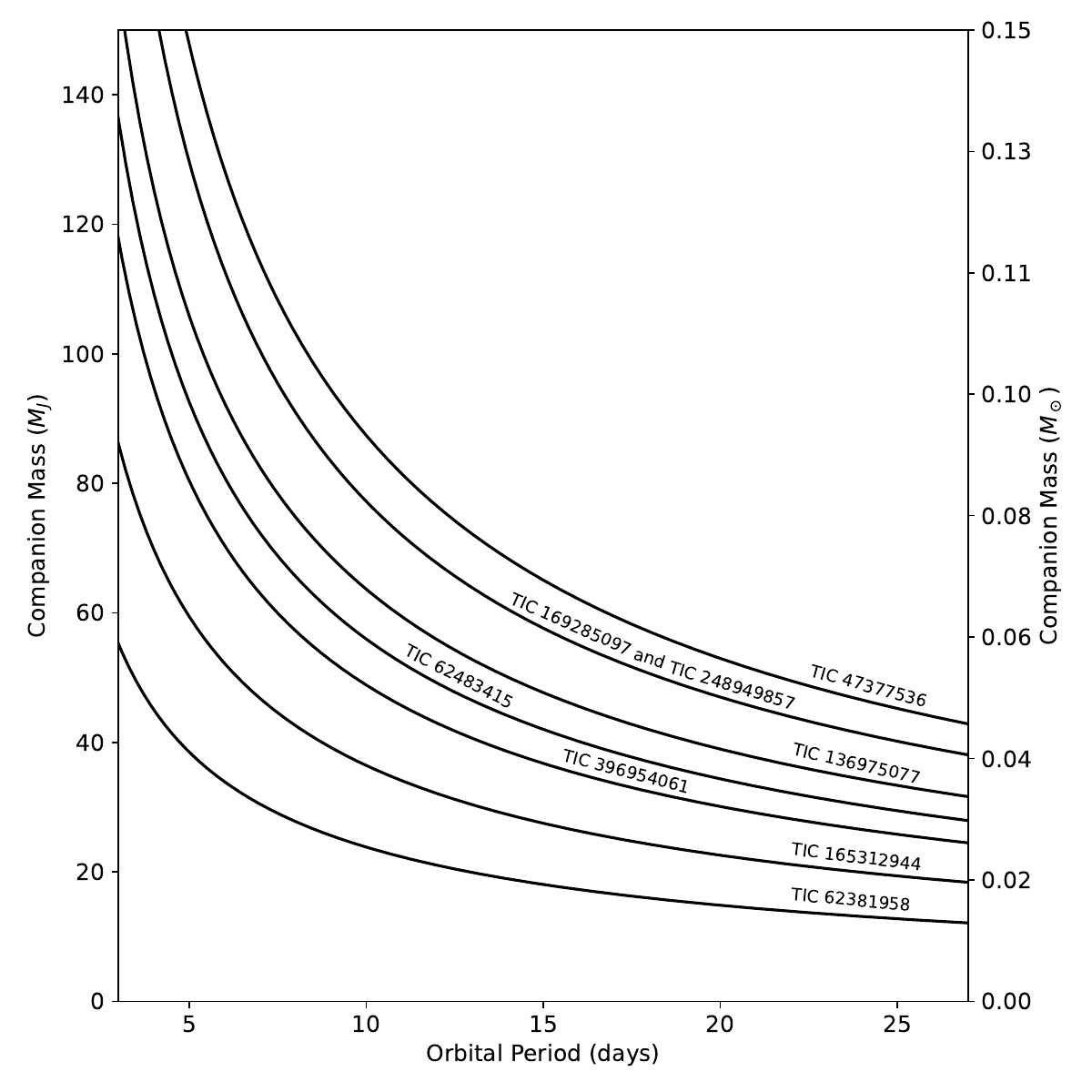} 
\caption{Upper companion mass limits for i=90$^\circ$ for each target using all available \tess\ sectors in the analysis.}  
\label{fig:rejection_ranges} 
\end{figure}

Solving this equation for $M_{comp}$, we estimated the upper limit for the companion mass (under the assumption that $i =90^\circ$) to be 222.9 $M_J$ for orbital period $P_{orb}$ = 3 days and 42.8 $M_J$ for $P_{orb}$ = 27 days. (see Figure\,\ref{fig:rejection_ranges}). 
\\



TIC\,62381958 (EC\,01541-1409, GD\,1053, Gaia\,DR2\,5149241067178231552) was found to be a pulsating hot subdwarf by \citet{kilkenny09}, who reported six p-mode frequencies. Later, \citet{reed12} detected $\sim$\,30 pulsation frequencies including all those reported by \citet{kilkenny09}. \citet{reed12} also reported that there are no indications that this is a binary system from the spectroscopic observations. \citet{baran23} used Sector\,30 \tess\ data and detected 38 frequencies including both p- and g-modes. Using our analysis approach, we found f$_6$, f$_7$ and f$_{30}$ to be significant in individual data subsets and adopted these frequencies for the pulsation timing analysis (see Figure\,\ref{fig:OC1}).  We found no variation with a light time delay variation amplitude $\geq$ 1.59\,sec. Only the light time delay data for f$_7$ were used since it provides the lowest average noise level. We estimated an upper companion mass limit for $i =90^\circ$ to be 55.1 $M_J$ for $P_{orb}$ = 3 days and 12.1 $M_J$ for $P_{orb}$ = 27 days. (see Figure\,\ref{fig:rejection_ranges}).
\\

TIC\,62483415 (EC\,22221-3152, Gaia\,DR2\,6601695863046409600) was found as a variable sdB star by \citet{kilkenny09}. \citet{barlow17} derived T$_{\rm eff}$\,=\,35\,600$\,\pm\,$600\,K, $\log\,$g\,=\,5.86\,$\pm$\,0.15, and $\log$ N(He)/N(H)\,=\,$-1.4 \pm 0.3$. \citet{baran23} detected twelve pulsation frequencies using Sector\,28 \tess\ USC data. We analyzed the three frequencies with the highest amplitudes reported by \citet{baran23}, f$_1$, f$_2$, and f$_6$, that are shown in Figure\,\ref{fig:OC1}. Our results show no variation with amplitude higher than 3.59\,sec.  Only light time delay data for f$_1$ was used since it provides the lowest average noise level. 
We estimated an upper companion mass limit for $i =90^\circ$ to be 136.1 $M_J$ for $P_{orb}$ = 3 days and 27.9 $M_J$ for $P_{orb}$ = 27 days. (see Figure\,\ref{fig:rejection_ranges}).
\\

\begin{figure*}[] 
\centering
\includegraphics[width=\hsize]{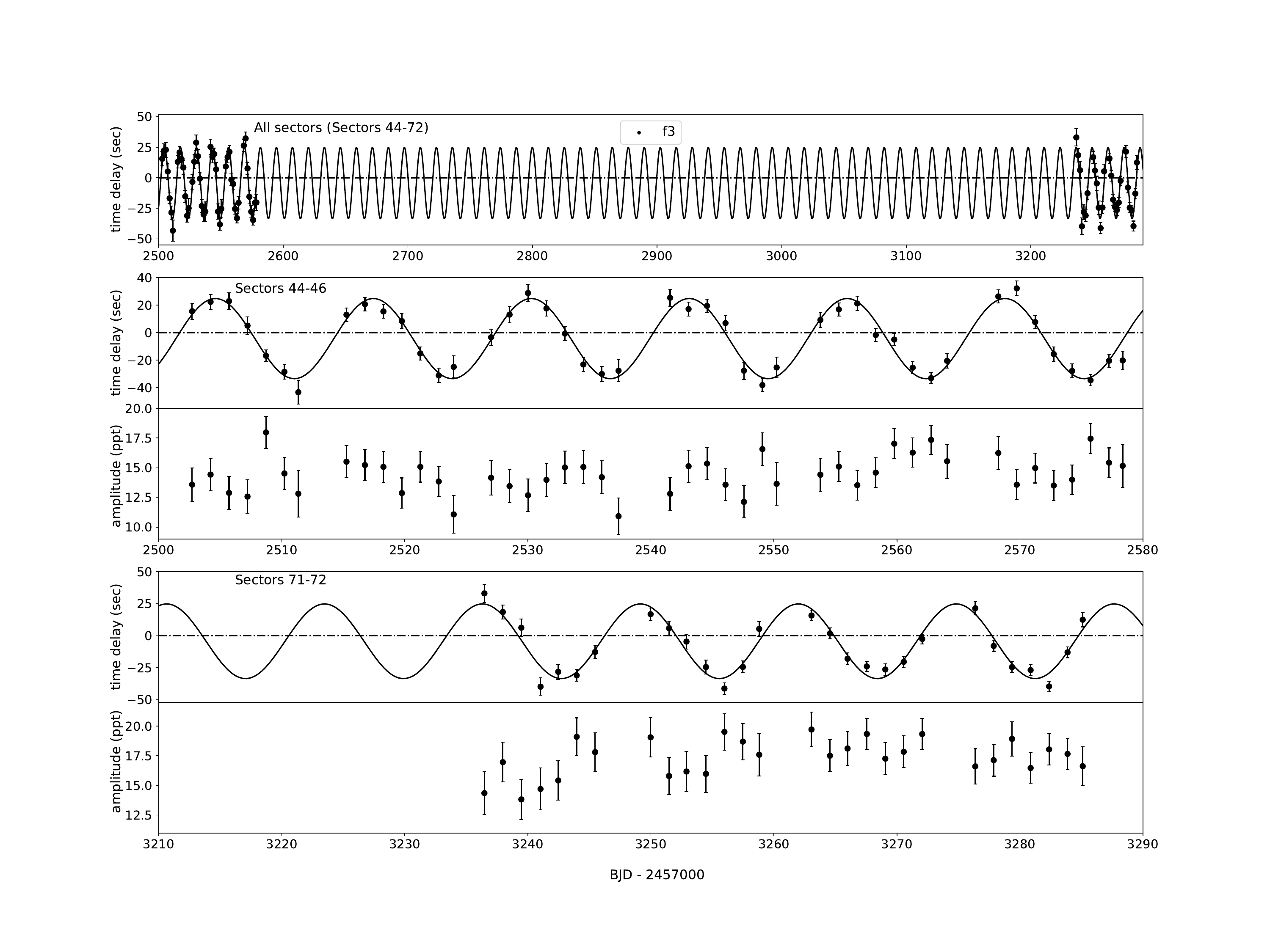}
\caption{Pulsation timing analysis of TIC\,69298924 using f$_3$ (the largest amplitude pulsation). ($Row 1$) Pulsation timing analysis of all available sectors, Sectors 44-46, and 71-72. ($Row~2~and~3:~Top$) The pulsation timing analysis of Sectors 44-46 and Sectors 71-72. The solid curve shows the best-fit sinusoid.  ($Row~2~and~3:~Bottom$) The pulsation amplitude variations of each sector.} 
\label{fig:OC_variation_19}
\end{figure*}

TIC\,69298924 (J08069+1527, Gaia\,DR2\,654866823401111168, GALEX\,J080656.7+152718) was classified as an sdB star by \citet{vennes11}. \citet{baran11} found it to be a hybrid pulsator and detected two g-mode and two p-mode frequencies. \citet{baran23} analyzed Sector\,44-46 \tess\ USC data and detected only four frequencies in the p-mode region. They suggested that f$_2$, f$_3$, and f$_4$ may be a rotationally split triplet. Assuming this interpretation is correct the rotation period would equal 12.9\,d. We applied the pulsation timing analysis only to frequency f$_3$ since it was the only significant frequency in individual data subsets. The result of our analysis is shown in  Figure\,\ref{fig:OC_variation_19}. The solid line in Figure\,\ref{fig:OC_variation_19} is our fit to the data points. A variation with a period of 12.83569\,$\pm$\,0.00037\,d is noticeable, while we found no significant variation of the amplitude of f$_3$.

Assuming the light time delay variation is caused by an unseen companion, the mass function $f = 0.161\,\pm$\,0.022\,M$_\odot$. The estimated amplitude of the RV variation is K$_{\rm sdB}$\,=\,49.4\,$\pm$\,2.2\,km\,s$^{-1}$. The existence of the additional companion could be verified by radial velocity observations. Following Equation\,28 of \citet{shibahashi12}, we estimated the sum of amplitudes of side frequencies f$_2$ and f$_4$ to be 7.03(14)\,ppt. According to \citet{baran23}, the sum of the observed amplitudes of these side frequencies are 7.18(42)\,ppt. These two values agree within their uncertainties. Given both the lack of the amplitude variation of f$_3$ and the consistent sum of observed/predicted amplitudes of side frequencies in this triplet, we conclude that the light time delay variation is caused by an orbiting companion to this pulsating sdB star, making the object a binary system. The light time delay variation amplitude is \,29.1\,$\pm$\,1.3 seconds, and the corresponding semi-major axis a$_{\rm sdB}\,\sin{i}$\,=\,0.0583\,$\pm$\,0.0026 $au$. The orbital solution is presented in Table\,2.

\begin{figure}[] 
\plotone{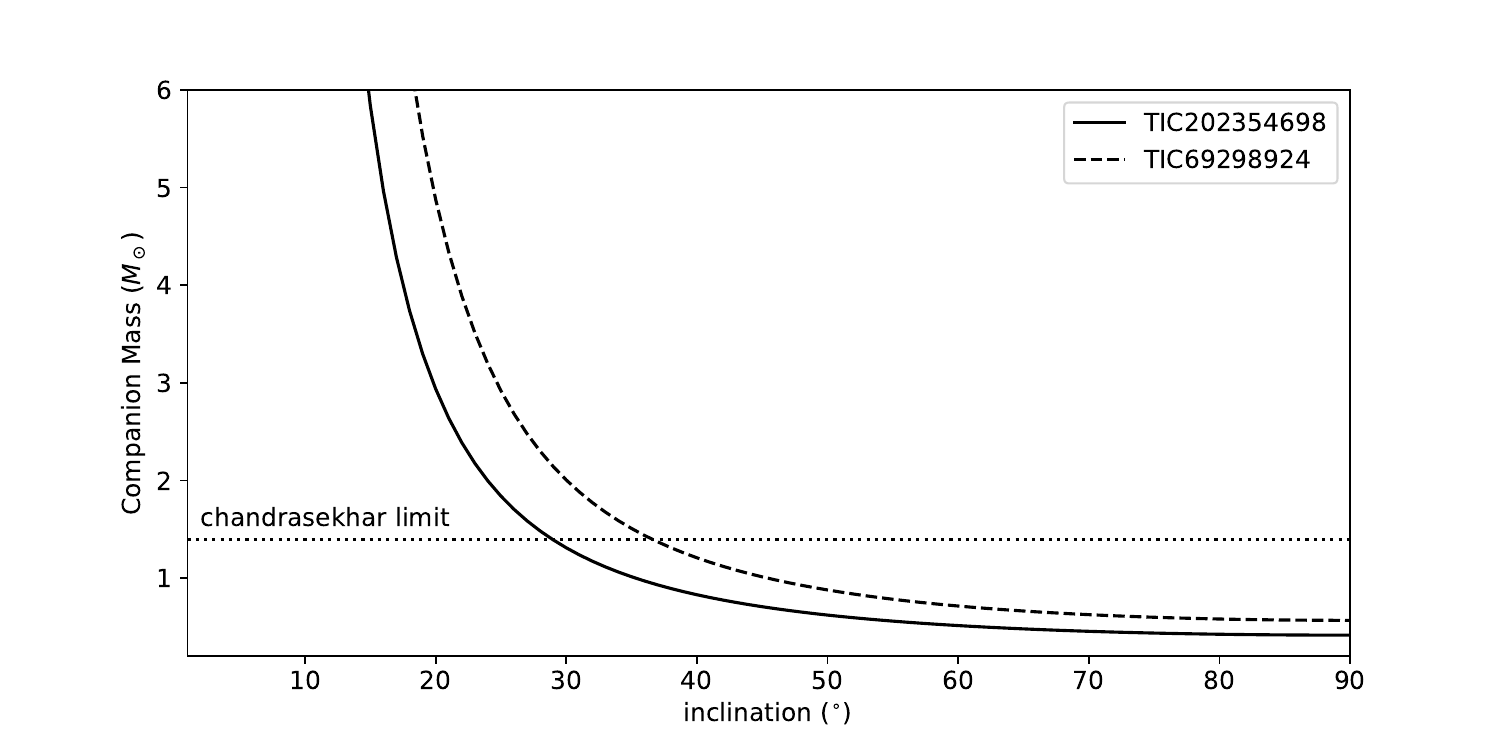} 
\caption{Possible companion masses for TIC\,202354698 and TIC\,69298924 as a function of orbital inclination.} 
\label{fig:OC_compmass} 
\end{figure}

\begin{figure}[] 
\plotone{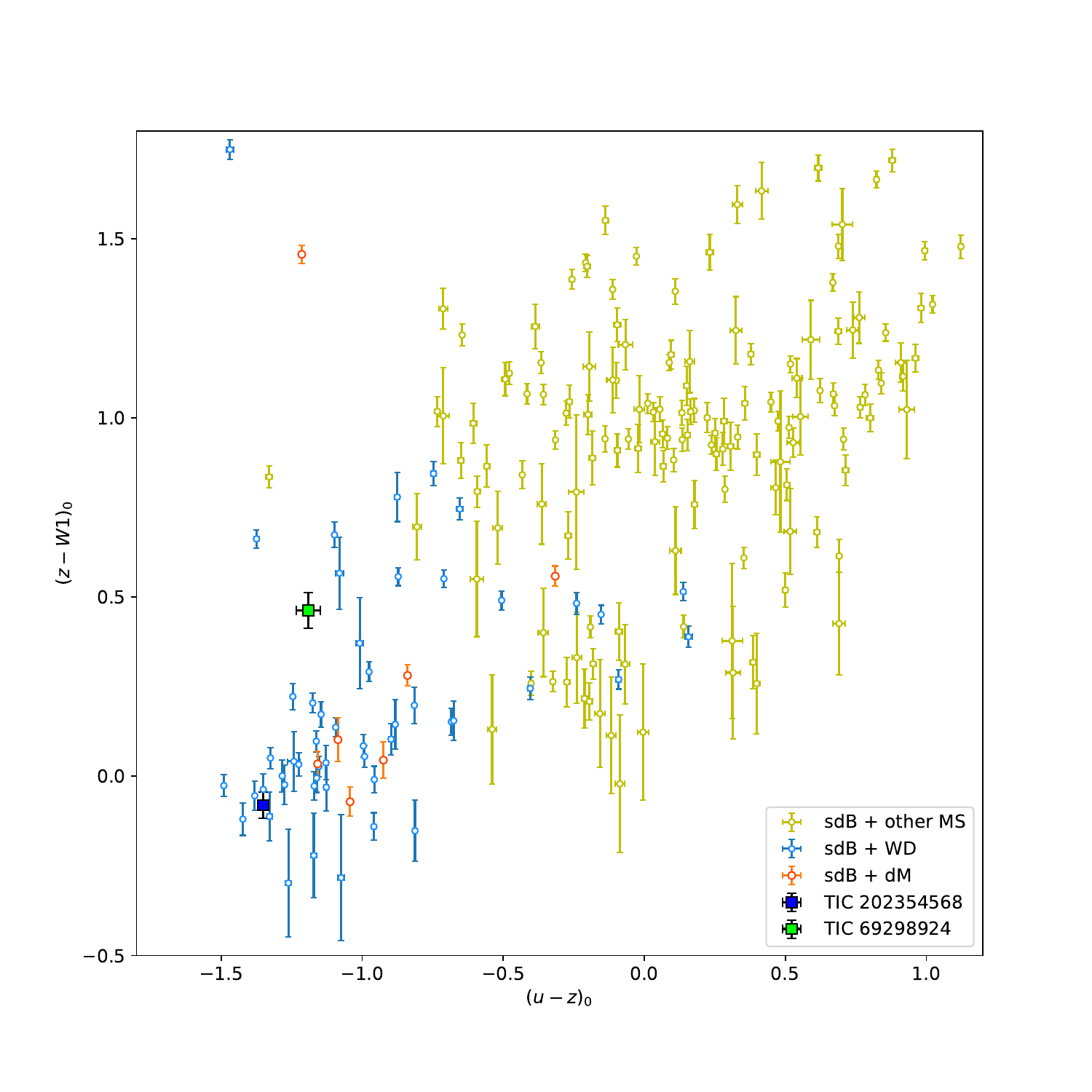} 
\caption{Color-color diagram for all available sdB binaries} in the \citep{geier20} database. TIC\,202354698 and TIC\,69298924 are indicated by filled squares.
\label{fig:color_diagram} 
\end{figure}

According to the BPS models, short orbital period sdB binaries have either MS or WD companions. Both Figure\,3 of \citet{ge22} and Table\,9 of \citet{baran19} show that sdB+MS systems have orbital periods shorter than 1.25\,d. This indicates that this companion is likely a WD. According to Figure\,\ref{fig:OC_compmass} the minimum mass of a companion is 0.568\,M$_\odot$. Note that for the WD interpretation to be valid, the orbital inclination cannot have $i < 36^{\circ}$ or the unseen companion would exceed the Chandrasekhar limit.  
Figure\,\ref{fig:color_diagram} shows the color–color diagram of known sdB binary systems. The position of the star in this diagram also supports the WD interpretation of the companion to the sdB star.
\\

TIC\,136975077 (V2203\,Cyg, KPD\,2109+4401, Gaia\,DR2\,1970289725639077120) was classified as an sdB star by \citet{downes86}. \citet{koen98} isolated seven p-mode frequencies. \citet{heber00} performed NLTE and LTE model atmospheres using Keck HIRES spectra and obtained T$_{\rm eff}$\,=\,31\,800\,$\pm$\,600\,K, $\log$\,g\,=\,5.76\,$\pm$\,0.05, and $\log$(He/H)\,=\,$-2.23 \pm 0.10.$ \citet{jeffery00} obtained radial velocities of 2.49 and 2.68\,km\,s$^{-1}$ at frequencies 5.09 and 5.48\,mHz and concluded that they cannot rule out the possibility that this star has a companion. \citet{baran24} used Sector\,55 and 56 \tess\ USC data and reported seven frequencies. We detected only f$_4$ to be significant in individual data subsets. The pulsation timing analysis using f$_4$ is shown in Figure\,\ref{fig:OC1}. We found no variations with amplitude higher than 4.05\,sec.  From this value, we estimated an upper companion mass limit for $i =90^\circ$ to be 156.6 $M_J$ for $P_{orb}$ = 3 days and 31.6 $M_J$ for $P_{orb}$ = 27 days. (see Figure\,\ref{fig:rejection_ranges}).
\\

TIC\,165312944 (PG\,1219+534, KY\,UMa, Gaia\,DR2 1572188107440718848) was classified as a variable sdB star by \citet{koen99}. They also indicate that its spectra and photometric colors are consistent with a single star, and there is no sign of a companion.  According to the astroseismic analyses reported by \citet{charpinet05}, its mass is M$_{\rm sdB}$\,=\,0.457\,$\pm$\,0.012\,M$_{\odot}$ and R$_{\rm sdB}$\,=\,0.1397\,$\pm$\,0.0028\,R$_{\odot}$. Analyzing combined \tess\ USC data collected during Sectors\,48 and 49 \citet{baran24} detected nine pulsation frequencies. We used only the two frequencies with the highest amplitudes reported by \citet{baran24}, f$_6$ and f$_4$, for the pulsation timing analysis (see Figure\,\ref{fig:OC1}). We found no variations with amplitude higher than 2.39\,sec, 
from which we estimated an upper companion mass limit for $i =90^\circ$ to be 86.0 $M_J$ for $P_{orb}$ = 3 days and 18.4 $M_J$ for $P_{orb}$ = 27 days. (see Figure\,\ref{fig:rejection_ranges}).
\\
 
TIC\,169285097 (HE\,2341-3443, SB\,815, Gaia\,DR2\,2312392250224668288) was identified as an sdB star by \citet{graham73}, while \citet{ostensen10} found it to be a pulsator. \citet{kawka15} searched for binary systems using spectroscopy and radial velocity methods; however, they did not find any evidence of a companion. \citet{sahoo20} analyzed 120\,sec cadence photometry data of \tess\ Sector\,2 and obtained 43 pulsation frequencies, including six p-mode pulsations. \citet{baran23} used \tess\ USC data collected in Sector\,29 and reported 34 frequencies. In addition to Sector\,29 we also analyzed Sector\,69 USC data. Only two frequencies reported by \citet{baran23}, f$_{30}$ and f$_{31}$, were significant in individual data subsets, so those frequencies were used for the pulsation timing analysis (see Figure\,\ref{fig:OC1}). We found no variations with amplitude higher than 4.86\,sec, corresponding to an upper companion mass limit for $i =90^\circ$ of 195.3 $M_J$ for $P_{orb}$ = 3 days and 38.3 $M_J$ for $P_{orb}$ = 27 days. (see Figure\,\ref{fig:rejection_ranges}).
\\

TIC\,202354658 (PG\,1544+601, Gaia\,DR2\,1627040062490757120) was classified as an sdB star by \citet{green86}. \citet{baran24} analyzed consecutive \tess\ Sector\,48-51 USC data and detected eight pulsation frequencies identified as p-modes. Three frequencies, f$_5$, f$_6$ and f$_7$, indicate the possibility of being rotationally split modes. Assuming the interpretation of a triplet is correct, the rotation period would be 17.3\,d. \citet{baran24} found two other equally spaced frequency patterns among f$_1$, f$_2$, and f$_3$, and f$_4$, f$_5$, and f$_8$, with both spacing equal to 2.09\,$\upmu$Hz. The authors pointed out that this could be due to a companion with a period of 5.5\, d. To test this possibility we split \tess\ USC data into 0.75 d data subsets and performed our pulsating timing analysis.

\begin{figure*}[] 
\centering
\includegraphics[width=\hsize]{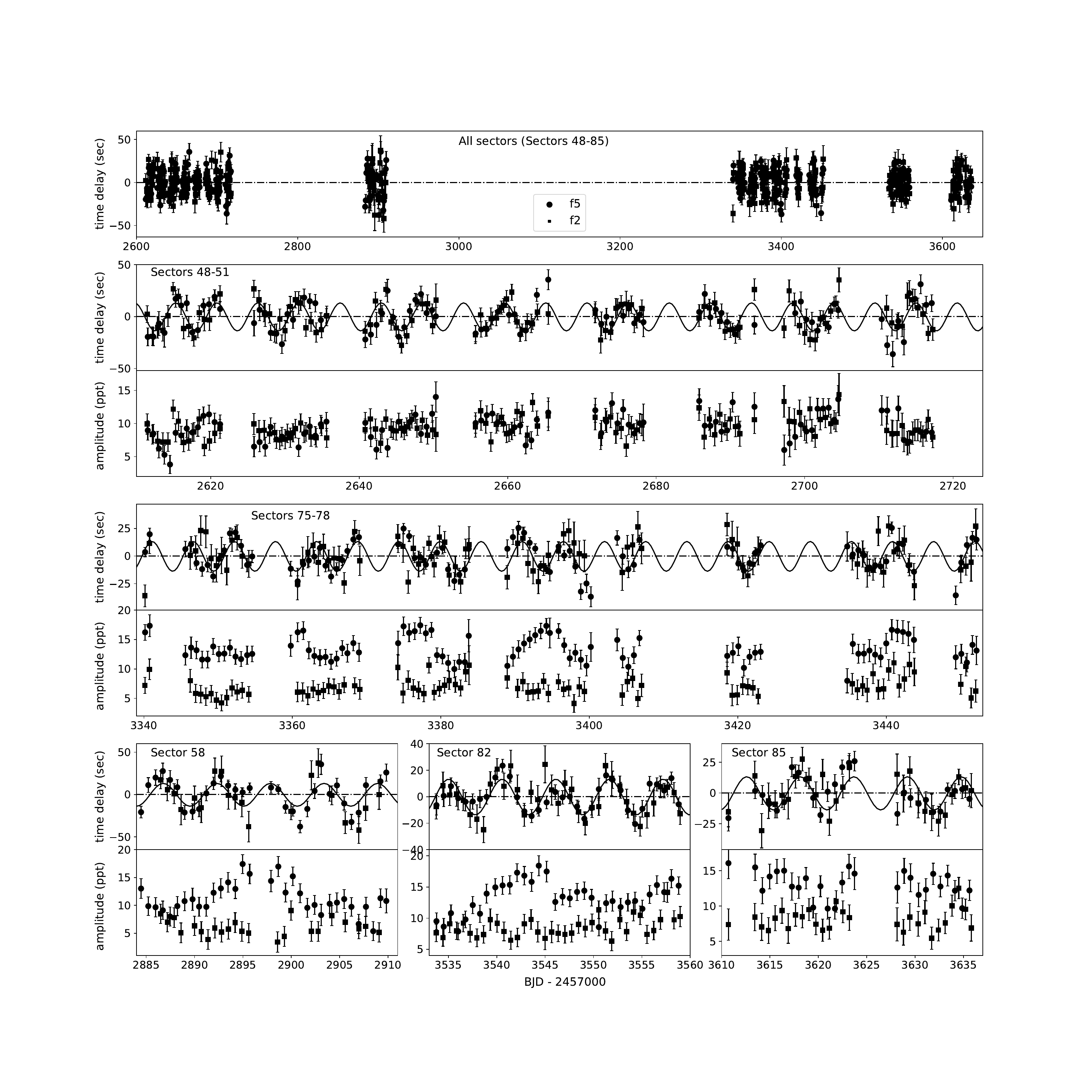}
\caption{Pulsation timing analysis of TIC\,202354658 using f$_5$ (the largest amplitude pulsation - shown in ``$\bullet$'') and f$_2$ (the second largest amplitude pulsation - shown in ``$\blacksquare$''). ($Row 1$) Pulsation timing analysis of all available sectors, Sectors 48-51, 58, 75-78, 82, and 85. ($Row 2-4: Top$) The pulsation timing analysis of each sector. The solid curve shows the best fit sinusoid. ($Row 2-4: Bottom$) The pulsation amplitude variations of each sector. } 
\label{fig:OC_variation_11}
\end{figure*}

For the pulsation timing analysis we used only two frequencies that were significant in individual data subsets of data, {\it i.e.} f$_2$ and f$_5$. The result of our analysis is shown in Figure\,\ref{fig:OC_variation_11}. The solid line in Figure\,\ref{fig:OC_variation_11} is our fit to the light time delay data points. It represents a sinusoidal variation with a period of 5.54078\,$\pm$\,0.00061\, d and and amplitude of 13.49\,$\pm$\,0.65\,sec. Based on the orbital period, similarly to TIC\,69298924, the companion of TIC\,202354658 is likely a WD. The amplitudes of both f$_2$ and f$_5$ show no significant variation. The minimum mass of a companion is 0.416\,M$_\odot$ (see Figure\,\ref{fig:OC_compmass}). Note that if the inclination is below $i$\,=\,29$^{\circ}$, the total mass of the system would exceed the Chandrasekhar limit. Figure\,\ref{fig:color_diagram} shows the color–color diagram of known sdB binary systems. The position of the star in this diagram is also consistent with the WD interpretation. Since the rotation period of the sdB star is 17.3 d, which is longer than the orbital period of 5.54 d,  this binary system is subsynchronous.

The derived sums of amplitudes of side frequencies f$_1$ and f$_3$, as well as f$_4$ and f$_8$, are 2.96(25)\,ppt and 2.98(25)\,ppt, respectively. According to \citet{baran24}, the sum of the observed amplitudes of these two pairs of side frequencies are 2.76(32)\,ppt and 2.56(33)\,ppt, respectively. The corresponding values agree within their uncertainties. The lack of amplitude variations of f$_2$ and f$_5$, as well as consistent sums of amplitudes of side frequencies indicates the presence of an unseen companion, supporting the interpretation of this object as a binary system. The orbital solution is included in Table\,\ref{tab:orbitalinfo}. The mass function as computed from the pulsation timing variation is f=0.086\,$\pm$\,0.012 M$_\odot$. The estimated amplitude of the RV variations is K$_{\rm sdB}$\,=\,53.1\,$\pm$\,2.6\,km\,s$^{-1}$.
\\

TIC\,248949857 (PB\,8783, EO\,Cet, Gaia\,DR2\,2482171590176492928) is a known subdwarf O (sdO) and main-sequence F spectroscopic binary system \citep{jeffery00}. The primary component was originally classified as an sdB pulsator by \citet{koen97}. \citet{ostensen12} analyzed the spectra and reclassified this as an sdO star. \citet{baran23} used Sector\,30 \tess\ USC data and confirmed 11 frequencies that were previously reported by \citet{odonoghue98} and \citet{vangrootel14}. We found frequencies f$_8$ and f$_9$ to be significant in individual data subsets, so those frequencies were used for the pulsation timing analysis (see Figure\,\ref{fig:OC1}). We detected no variation with amplitude higher than 4.84\,sec. Using this, we estimated an upper limit of companion mass for $i =90^\circ$ to be 193.8 $M_J$ for $P_{orb}$ = 3 days and 38.1 $M_J$ for $P_{orb}$ = 27 days. (see Figure\,\ref{fig:rejection_ranges}). We expect that the progenitor of this binary went through the RLOF and the orbital period is at least hundreds of days, therefore one sector of \tess\ data (about 27\,d) is insufficient to detect light time delay variations.
\\

TIC\,396954061 (2M\,0415+0154, Gaia\,DR2\,3259060049366022400) is a known sdB pulsator. \citet{oreiro09} detected three p-mode frequencies and derived T$_{\rm eff}$\,=\,34\,000 $\pm$\,500\,K and $\log$\,g\,=\,5.80\,$\pm$\,0.05. Their spectra and radial velocity search concluded that this is either a single or an sdB+WD system with a period longer than a few days. \citet{baran23} used Sector\,32 \tess\ USC data and found three pulsation frequencies, one of which is the same frequency reported by \citet{oreiro09}. We found that only f$_3$ is significant in individual data subsets, so we only used f$_3$ for the pulsation timing analysis (see Figure\,\ref{fig:OC1}).  We detected no variation with amplitude higher than 3.16\,sec. Using this, we estimated an upper companion mass limit for $i =90^\circ$ to be 117.5 $M_J$ for $P_{orb}$ = 3 days and 24.5 $M_J$ for $P_{orb}$ = 27 days. (see Figure\,\ref{fig:rejection_ranges}).
\\

\begin{deluxetable}{ccc}[h]
\tablecaption{Orbital parameters for TIC 202354698 and TIC 69298924 binary systems}
\label{tab:orbitalinfo}
\tablehead{\colhead{Parameters} & \colhead{TIC 202354698} & \colhead{TIC 69298924}} 
\startdata
Period, P (d) & 5.54078 $\pm$  0.00061 & 12.83569 $\pm$  0.00037 \\
Amplitude, a$_{\rm sdB} \sin{i}$ (seconds) & 13.49 $\pm$ 0.65 & 29.1 $\pm$ 1.3\\
Mass function, f (M$_{\odot}$) & 0.086 $\pm$ 0.012 & 0.161 $\pm$ 0.022\\
Radial velocity of sdB star, K$_{\rm sdB}$ (km/s) & 53.1 $\pm$ 2.6 & 49.4 $\pm$ 2.2\\
\enddata
\end{deluxetable}

\section{Conclusions}\label{sec:conc}
We used Transiting Exoplanet Survey Satellite ultra-short cadence data to search for companions to pulsating sdB stars with the light time delay. We focused on orbital periods shorter than 13.5\,d which are detectable with one sector of \tess\ data (about 27\,d). We investigated ten pulsating sdB stars reported by \citet{baran23,baran24} and detected clear indications of pulsation modulation characteristic of binary systems in TIC\,202354658 and TIC\,69298924. For the TIC 202354658 system, we find that the orbital period is 5.54078\,$\pm$\,0.00061\,d and a$_{\rm sdB}\,\sin{i}$ is 13.49\,$\pm$\,0.65\,sec. If the orbital inclination is below $i$\,=\,29$^{\circ}$, the mass of this unseen companion exceeds the Chandrasekhar limit. Since the stellar rotation period of TIC\,202354658 is 17.3 d, this binary system is an example of subsynchronous system. For the TIC\,69298924 system, the orbital period is 12.83569\,$\pm$\,0.0003\,d and a$_{\rm sdB}\,\sin{i}$ is 29.1\,$\pm$\,1.3\,sec. If the inclination is below $i$\,=\,36$^{\circ}$, the mass of this unseen companion exceeds the Chandrasekhar limit. Based on the orbital period length and a color-color diagram, we concluded that in both cases the companions are likely to be white dwarfs. We also estimated orbital radial velocity amplitudes as an aid in planning future spectroscopic observations that can be used to test our analysis and conclusions.

Although many observational studies have been done on the short orbital period sdB systems formed by the common envelope channel, more than half of them have orbital periods shorter than 0.6\,d and the study of sdB binaries with orbital periods longer than 5\,d is limited. Searching for binaries with orbital periods longer than typical CE binary systems, i.e.\ longer than a few days, provides valuable insights into sdB binary formation theories.

\newpage
\acknowledgments

We acknowledge research support from the National Science Foundation (NSF) under Grant No. AST-2108975. TO acknowledges research support from the National Aeronautics and Space Administration (NASA) under Grant No. 80NSSC24K0494. \\

This paper includes data collected by the Transiting Exoplanet Survey Satellite \tess\ mission. Funding for the \tess\ mission is provided by the NASA Explorer Program.\\

All the TESS data used in this paper can be found in MAST: \cite{tess-spoc} (DOI: \href{https://doi.org/10.17909/t9-wpz1-8s54}{10.17909/t9-wpz1-8s54}).

\vspace{5mm}
\facilities{TESS}


\bibliographystyle{aasjournal}
\expandafter\ifx\csname natexlab\endcsname\relax\def\natexlab#1{#1}\fi
\providecommand{\url}[1]{\href{#1}{#1}}
\providecommand{\dodoi}[1]{doi:~\href{http://doi.org/#1}{\nolinkurl{#1}}}
\providecommand{\doeprint}[1]{\href{http://ascl.net/#1}{\nolinkurl{http://ascl.net/#1}}}
\providecommand{\doarXiv}[1]{\href{https://arxiv.org/abs/#1}{\nolinkurl{https://arxiv.org/abs/#1}}}
\bibliography{myrefs}



\end{document}